\documentclass{egpubl}
\usepackage{EGUKCGVC2024}
\usepackage[pass]{geometry}
\usepackage{cancel}
 
\WsPaper           

\usepackage[T1]{fontenc}
\usepackage{dfadobe}  

\usepackage{cite}  
\BibtexOrBiblatex
\electronicVersion
\PrintedOrElectronic
\ifpdf \usepackage[pdftex]{graphicx} \pdfcompresslevel=9
\else \usepackage[dvips]{graphicx} \fi

\usepackage{egweblnk} 

\title[Visual Storytelling: A Methodological Approach]%
      {Visual Storytelling: A Methodological Approach to Designing and Implementing a Visualisation Poster}

\author[Owen and Roberts ]
{\parbox{\textwidth}{\centering 
R.S.Owen\thanks{r.s.owen@bangor.ac.uk}\orcid{XXXX} and
J. C. Roberts\thanks{j.c.roberts@bangor.ac.uk} \orcid{0000-0001-7718-3181}
}
        \\
{\parbox{\textwidth}{\centering Bangor University, UK\\
       }
}
}

\renewcommand*{\backref}[1]{
}

\begin{document}


\maketitle
\begin{abstract}
   We present a design study of developing a visualisation poster. Posters can be difficult to create, and the story on a poster is not always clear. Using a case-study approach we propose three important aspects: the poster should have a clear focus (especially a hero visualisation), envisioning its use helps to drive the important aspects, and third the essence (its fundamental concept and guiding idea) must be clear. We will use case studies that have focused on the use of the Five Design-Sheet method (FdS) as a way to sketch and plan a visualisation, before successfully implementing and creating the visual poster. The case studies serve as a practical illustration of the workflow, offering a means to explain the three key processes involved: (1) comprehending the data, (2) employing a design study with the FdS (Five Design-Sheet), (3) crafting, evaluating and refining the visualisation.
\end{abstract}  
\section{Introduction}
We are witnessing a \textit{``visualisation of culture''}~\cite{Beer2013}, where data and visualisations play a central role, and society collectively engages in discussions about them. Consequently, creating data-driven stories and designing impactful posters should be essential objectives of the developer. Visualisation experts claim that by creating accessible and transparent visualisations we can offer a greater understanding of data~\cite{Nino2008} promoting the goal `lets do good with data together'~\cite{Periscopic2020}. Making data readable and accessible is crucial, and we must move away from the stereotype that \textit{``scientists are seen as intelligent and hard-working but not interested in people''}\cite{Beardslee1961}. Computer scientists have been generalised as being less creative as they are keen to avoid romanticism and obscurantism, and \textit{``define creativity in terms of novel combinations of familiar ideas''}~\cite{Boden1994}. They often wish to create posters of their work, but the posters can be difficult to comprehend, and the story not clear. Individuals must consider the purpose, audience, data, content, design and layout of their work.
Additionally, the choice of software and specific display requirements for the poster must be considered.

We explore the requirements of creating an effective visualisation poster and the methodological approaches needed. This paper works towards developing a methodology for storytelling and poster visualisation creation. We present our argument and a case study of examples, and discuss the idea, working towards a fuller theory of design, and as such is part of ongoing wider research.

We follow three case studies, where developers create posters. They each use the \textit{Five Design-Sheet} (FdS)~\cite{RobertsHeadleandRitsos16} method, as a way to explore the ideas. While each are novices (having only presented a few posters each) they have all studied visualisation and understand basic design strategies.
E.g., they know about ``overview first, zoom and filter, and details on demand''~\cite{Shneiderman1996}, and the importance of gaining an overview and identifying different patterns~\cite{Keim2002}.

\begin{figure*}[t]
    \centering
    \includegraphics[width=\textwidth]{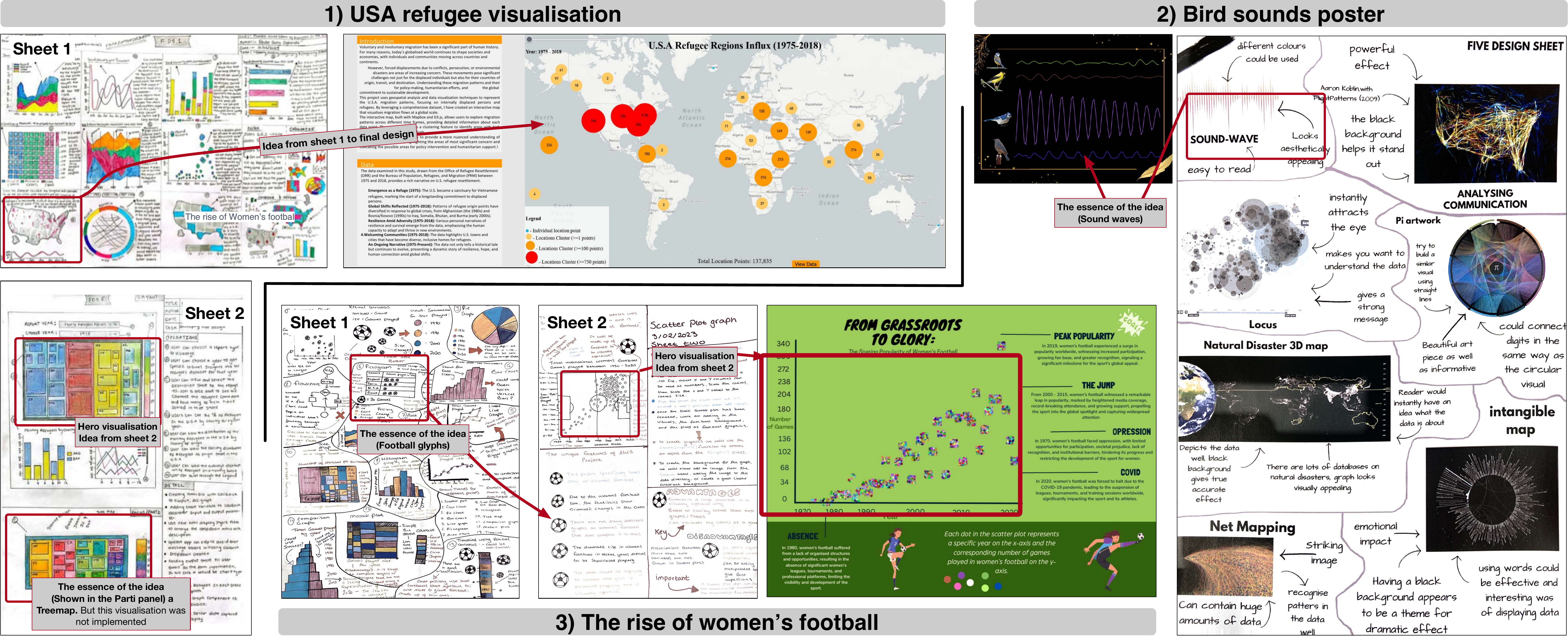}
    \caption{The three case studies. 1) USA refugee data. Showing the first sheet, sheet~2 and the poster. It demonstrates a range of alternative solutions (sheet~1) and a proposed essence idea (the Treemap visualisation), although the Treemap design of the sheet~2 was not chosen for the final poster; instead one of the earlier designs (the map visualisation) was chosen to be the main hero visualisation. Study 2) Bird sounds poster. Showing the first sheet, and a principal idea (the sound waves). Case study 3 `The rise of Women's football'' (bottom). The final poster (in green) shows the main hero visualisation. The idea essence is to encode the data into a football shaped glyph (as highlighted from Sheet~1 and in the Parti panel (bottom left) of sheet~2. The poster presents information of the rise of women's football from 1970 to 2021. 
    }
    \label{fig:caseStudy}
\end{figure*}

\section{Ideology and related work}
Effective story-telling \textit{``requires skills like those familiar to movie directors, beyond a technical expert's knowledge of computer engineering and science''}~\cite{Gershon2001}, however, traditional storytelling typically presents a sequence of events, whereas data stories can be arranged in a linear sequence, encouraging verification, explanations and can also be interactive~\cite{Segel2010}. Creating visualisation stories on posters is a valuable skill, but it can be challenging for novices. In a visualisation course at the higher education level, developing a poster can serve as an effective and meaningful assignment. We focus on creating visualisation posters in the Creative Visualisation and Information Visualisation modules at the university. These assignments are given to third-year students pursuing their BSc in Computing. Furthermore, PhD students are encouraged to create posters of their work, at an early stage of their training. We propose three primary concerns that a developer should consider when designing and creating a visualisation poster.

\textbf{When making an appropriate public facing poster, it is not only a matter of displaying the visualisations along with text, but that a strong focus is required.} We emphasise that a `hero visualisation' is required.  If every visualisation is the same size then people do not know where to look. We define a `hero visualisation' as a standout or central visual representation that effectively conveys key insights from complex data. Associated labels, other visualisations and text can be added alongside, but the key point is that it needs to have a central focus. The aim of the poster creator is to capture the audience's attention and drive the main narrative. We encourage that these visualisation posters are designed to be highly engaging, clear, and impactful, and serve to become a focal point in presentation. The hero visualisation, as the central focus, is crucial because it represents the main results and should take center stage. Individuals should highlight what they have spent the most time on, such as their data analysis and visualisation; these elements should be the primary focus. While words, text, and legends are important for confirming and clarifying information, they should not dominate the viewer's attention. 

\textbf{Second, that understanding its use and location is crucial.} The creator should imagine themselves presenting the poster at the event and within the specific location where it will be displayed. Visualising this scenario helps the creator focus on the key elements that will capture the audience's attention and effectively convey the main message. By considering the environment and context in which the poster will be used, the creator can better prioritise information, ensuring that the most important points are clear and engaging for viewers. After all, the creator will often be present with their poster, using it as a visual aid. 
This approach not only aids in refining the content but also enhances the overall impact and usability of the poster.

\textbf{Third, the poster (and by extension, the creator) must emphasise the main message of the data and/or data visualisation.} It is crucial for the creator to grasp the core essence and ensure the poster effectively communicates it. We express the idea of the `essence' of the work (a concept that is named the Parti, in the FdS. In architecture the `parti' refers to the fundamental concept or guiding idea that shapes the overall design, which serves as the underlying theme that influences every aspect of the architectural plan. This idea can be used in visualisation, especially poster design. 

Consequently, we propose that using the FdS enables developers to fulfil each of these goals; to consider alternative visualisations and create a design with a main hero image (through the many ideas on Sheet~1 of the FdS process). To imagine and confirm which idea works best, and would be suitable for use at the poster venue/conference (after considering the many ideas on Sheet~1 and the three alternatives on sheets 2,3,4). To find the essence (its Parti) on the second panel of each sheet 2,3,4 and 5~\cite{RobertsHeadleandRitsos16}.


\section{Case Studies Overview}
In this section, we provide a brief overview of the three case studies that form the basis of our design study. Each case study involves a third-year student pursuing their BSc in Computing with some experience in data visualisation. Each student, despite being relatively inexperienced, demonstrated the potential of the FdS in guiding their design process, enabling them to create visually appealing and informative posters.

\noindent\textbf{Student 1 focused on visualising refugee data.} The data concerns internally displaced persons and refugees in the USA (see figure \ref{fig:caseStudy} - top left corner of the image). The objective was to create a comprehensive geospatial analysis that highlights migration patterns within the country. The project utilised \textit{Mapbox} and \textit{D3.js} to develop an interactive map that not only visualises migration flows but also includes a clustering feature to identify areas with high concentrations of displaced individuals. This visualisation aimed to inform policy interventions and humanitarian support by making the data more accessible and understandable.

\noindent\textbf{Student 2's project centered on visualising bird songs} using a dataset titled ``Sounds Of 114 Species Of Birds''\cite{Mohanty2022}  (see Figure~\ref{fig:caseStudy} -- top right corner of the image). The goal was to represent bird songs as visually engaging sound waves, categorised by species. Inspired by artists like William Shaw \cite{Shaw} and Nathalie Miebach \cite{Miebach}, Student 2 aimed to combine scientific data with artistic expression, thereby highlighting the unique acoustic signatures of different bird species. The project employed \textit{Processing} to create these vibrant visualisations, offering an abstract way to appreciate bioacoustic data.

\noindent\textbf{Student 3 explored the historical growth of women's football} from 1970 to 2021 (see Figure~\ref{fig:caseStudy} -- bottom on the image) through a scatter plot diagram. The project aimed to visualise the rise in popularity and participation in women's football over time. By following Andy Kirk’s design principles from \textit{``Data Visualization: A Handbook for Data Driven Design''}~\cite{Kirk2019}. Student 3 created a compelling scatter plot that effectively communicated the trends and milestones in the development of women's football.

\section{Sheet 1 of the FdS -- Investigating alternative ideas}

Sheet 1 of the FdS, also known as the ideas sheet, is used by the designer to consider alternative ideas. They are encouraged to sketch and draw 10-15 potential ideas. When developing a poster, there are two key considerations to address. First, determine the story and how it presents the data. Second, focus on the layout of the poster itself.
Whilst brain-storming, the developer needs to sketch different ideas, and explore alternative and possible ways to display the data. These are shown by each of the students' work in Figure~\ref{fig:caseStudy}. E.g., Student 3 included: A scatter plot, flow chart, pie chart, bar chart, line graph, pictogram, area chart, a line graph with multiple data points, histogram, tree map, comparison graph and a mosaic plot. These graphs have been sketched as they all have the possibility to fit the requirement of displaying the data well, although the FdS doesn't have to be hand drawn it is recommended. A study by Schwamborn suggests that individuals learn more effectively when they engage in suitable generative processing, such as drawing during the learning process~\cite{Schwamborn2010}, however preferred learning styles are complex~\cite{Abrahamson2020} and web based-drawing can often be preferred, this style can open up opportunities for different aesthetics, broadening perceptual experiences. During this design stage we are analysing and evaluating each design with a view to improving them~\cite{Paul2001}, critical thinking is crucial here and decisions to move forward with designs, and drop a few designs begin. For example while the tree-map is a favoured method for visualising hierarchical data, it has been associated with being difficult for users to track layouts and recognise patterns of data that change over time~\cite{Tu2007} making it unsuitable for certain projects. Student 1 drew a line graph as they are considered to be a good choice for visualising time series data, however, a scatter plot was also put into the mix as they are associated with displaying main trends~\cite{WangEtal2016}. By filtering through the generated idea's case study 1 can exclude designs such as the comparison graph and the flowchart by circling back to the question ``what is the purpose of this graph''? Following critical thinking guides such as Richard Paul’s \textit{``Miniature guide to critical thinking'',}~\cite{paul2009miniature} they were deemed unable to fulfil the brief or desired outcome. Reflection is also key during this phase, following guides such as figures in publications\cite{franzblau2012graphs} 
help with decision making by encouraging questioning such as ``does this graph deserve to be in the project''? If they do not enhance the data, make it more accessible or transparent, further focus on designs that do. 

\subsection{Refine by considering three principal alternative designs}

When refining the design using the FdS methodology, individuals must select three main idea's (from sheet 1). They need to consider the data-story and the layout of the poster, especially when deciding their hero image design. On each sheet they include a sketch of the poster (the big picture), the main essence of the idea is described in the information panel (top right of the middle sheets) and an explanation of the main essence (the Parti) of the idea: the unique selling point, a most critical concept. In addition, individuals must consider the core components of the design (middle right panel of the middle sheets), and positives and negatives (bottom right of the middle sheets). While well-crafted visualisations can greatly benefit people, poorly designed ones are frequently ineffective or can even be misleading~\cite{Munzner2006}, this phase requires focusing on refinement which has been a core focus in software engineering since its inception, becoming increasingly more important during the last decade~\cite{Hallstrom2009}. Roberts et al.~\cite{RobertsHeadleandRitsos16} suggest that it is often helpful for individuals to conceptualise their designs by first sketching what it will look like (placing this sketch in the Big Picture panel on the FdS, top left) and capturing the essence of the idea (in the Parti panel on the FdS, bottom left).

For example, during this refining stage, Case Study 3 noticed when removing the lines that connect the time plots of their line graph, it became a scatter plot displaying a time series as a set of points, however, using point representations can fail to show changes between points, which line graphs are able to do~\cite{WangEtal2016}, the aspect ratio of a line graph also influences the perception of trends~\cite{cleveland1993visualizing}. 
While using FdS case study 3 concluded that each chart has advantages and disadvantages pointed out in the bottom right section of sheets 2 (see Figure~\ref{fig:caseStudy}). Scatter plots are a simple but powerful way of visualising two-dimensional point data and can visually display data trends with correlations between the two dimensions~\cite{Mayorga2013}. Daniel Keim describes scatter plots as \textit{``the most powerful tools for data analysis''}~\cite{Keim2002}, however, he also reveals that today's scatter plots have a high tendency to overlap, causing the data values to appear obscure~\cite{Keim2002}. Case study 3 also studied the timeline because they can use retinal variables and are useful for displaying events over time, however, Bevan argues they are more often used for storytelling, not as tools for visual analysis~\cite{Bevan2013}. Timelines have the advantage of being adaptable, you can swap text for images, linear or horizontal, curved or without a line or just structured chronologically. The timeline offers an innovative way of representing and comparing data chronologically in a visual format, however, when refining during this phase and reflecting on its main purpose, it was concluded that its simple nature restricted its ability to represent a large amount of data. This led to case study 3 creating their final design sheet based on the scatter plot, it is a versatile graph, can also include retinal variables and can present large data in a creatively visual way. 

\subsection{Sheet 5 -- Realisation Design}
Design realisation process can have a positive impact on future design and products~\cite{Albano1990}, by this stage primary objectives and design faults have been addressed and the final design has been reached through filtering and continuous improvement to optimise the design realisation process, this is now planned out on the fifth and final page of the FdS sheet (see Figure~\ref{fig:caseStudy} --- sheet 2). Visualising large data sets are likely to produce a busy, overlapping image, too many pixels can become over plotted resulting in loosing the important data pattern~\cite{Bertini2006}. Finding interesting trends and patterns in visualisations such as a large scatter plot can be challenging and the issue increases with the size of the scatter plot~\cite{Shao2015}. During this final stage of design, utilising helpful tools such as the \textit{``California Critical Thinking Skills Test''}~\cite{Facione2013} or in the case of case study 3 \textit{``The Critical Design Survey (CDS)''}~\cite{Roberts2023} are beneficial as we can reflect back and understand further improvements. The design ranks high with its suitability for the user and task, however, could be improved in other area's such as having suitable output and view types. Making knowledge visible so that it can be easily accessed, valued and appreciated is a valuable concept that has been used in these case studies~\cite{Eppler2007}.

All three case studies utilised the \textit{Five Design-Sheet methodology} 
to create their visualisations, case study 1 created a geospatial analysis using data visualisation techniques for representing U.S.A. migration patterns, focusing on internally displaced persons and refugees. It was an interactive map, built with \textit{Mapbox} and \textit{D3.js} and visualises migration flows globally and includes a clustering feature to highlight areas with high concentrations of displaced individuals, aiming to inform policy intervention and humanitarian support. Case Study 2 created a bold visual representation of bird songs. Using the dataset \textit{``Sounds Of 114 Species Of Birds''}~\cite{Mohanty2022}, the project depicted bird songs as brightly coloured sound waves, categorised by species. Inspired by artists like William Shaw and Nathalie Miebach, the visualisation in Processing combined scientific data with artistic expression, highlighting the unique sounds of each species and offering a new way to appreciate bioacoustic data. Case study 3 developed a scatterplot diagram visualising the rise of women's football from 1970 to 2021. Following Andy Kirk's design principles from \textit{``Data Visualisation: A Handbook for Data Driven Design''}~\cite{Kirk2019} and reflecting on design choices using the \textit{critical design survey} (CDS), the project created a compelling visualisation. The final CDS identified areas for improvement, noting that adding historical context could enhance the diagram's effectiveness.

\section{Discussion and conclusion}
The final projects offered valuable insights into the different stages of creating effective visualisations and utilising the \textit{Five Design-Sheet} methodology as an asset. The visualisations presented compelling narratives across different domains, showcasing the power of structured design processes in uncovering patterns and trends within complex datasets. The \textit{Five Design-Sheet methodology} played a pivotal role in shaping each visualisation, guiding the decision-making process through its systematic stages: comprehending the data, employing thoughtful design strategies, crafting, evaluating, and refining the final product. This structured approach facilitated the creation of visualisations that are not only visually appealing but also rich in information and accessibility. Structured methodologies like the \textit{Five Design-Sheet} methodology can significantly enhance the clarity and effectiveness of visual representations and emphasise the importance of iterative design processes in visualisation~\cite{AIGNER2007401}. The success of the visualisations created by the case studies confirm the importance of combining methodological approaches with creative design. This combination provides a powerful framework for future work in the field of data visualisation. By integrating best practices in data comprehension, design thinking, and visualisation techniques, researchers and practitioners can develop posters that effectively communicate their findings and engage a broad audience.

Each of the case studies underscores our strategy and key points. The first step is to understand the data and identify what is important, then highlight this significance with a `hero' image. This approach focuses the work, enabling the developer to stand by their poster and highlight the main ideas, ensuring the viewer clearly understands what is important. Subsidiary information, legends, and secondary stories can then be added around this central `hero' visualisation. Our second point, maintaining focus on the poster's use, is equally important. Throughout the design process, the designer should continuously ask, ``is this suitable for use at venue X?'' ``Consider whether the viewers will understand it'', if it ``meets their expectations'', and ``if the quality is up to the standard expected by the audience''. This keeps focus of the design.  Third, the poster must tell a clear story. When the creator can summarise the work in a few words (aided by the Five Design-Sheets process), with a clear Parti panel and a summary panel containing a concise title and summary information, the poster should ideally reflect this clarity and focus.

Although still in the early stages, our research has shown potential in our three-part focus and its integration with the \textit{Five Design-Sheet methodology}. Typically, people hastily put together their posters, but we encourage a more thoughtful and concise planning approach to create a more effective visualisation poster. In our ongoing research, we plan to further refine our approach by exploring advanced visualisation techniques that can enhance the clarity and impact of our posters. One interest is exploring the application of machine learning algorithms to automate aspects of data visualisation, potentially streamlining the process while maintaining visual integrity. Ultimately, our goal is to develop a framework that produces visually compelling posters but also effectively communicates complex data to diverse audiences across various academic and professional settings.
\clearpage

\bibliographystyle{eg-alpha-doi} 
\bibliography{eg_poster_BIB}       

\newcommand{\etalchar}[1]{$^{#1}$}
\begin{thebibliography}{\uppercase{AMM{\etalchar{*}}07}}

\bibitem[AA20]{Abrahamson2020}
\textsc{Abrahamson D., Abdu R.}:
\newblock {Towards an Ecological-dynamics Design Framework For Embodied-interaction Conceptual Learning: The Case of Dynamic Mathematics Environments}.
\newblock \emph{Educational Technology Research and Development 69}, 4 (2020), 1889--1923.
\newblock \href {https://doi.org/10.1007/s11423-020-09805-1} {\path{doi:10.1007/s11423-020-09805-1}}.

\bibitem[AK90]{Albano1990}
\textsc{Albano R., Keska J.}:
\newblock {Is Design Realization a Process? A Case Study}.
\newblock \emph{IEEE Transactions on Components, Hybrids, and Manufacturing Technology 13}, 3 (1990), 509--515.
\newblock \href {https://doi.org/10.1109/33.58852} {\path{doi:10.1109/33.58852}}.

\bibitem[AMM{\etalchar{*}}07]{AIGNER2007401}
\textsc{Aigner W., Miksch S., Müller W., Schumann H., Tominski C.}:
\newblock {Visualizing Time-oriented Data: A Systematic View}.
\newblock \emph{Computers \& Graphics 31}, 3 (2007), 401--409.
\newblock \href {https://doi.org/https://doi.org/10.1016/j.cag.2007.01.030} {\path{doi:https://doi.org/10.1016/j.cag.2007.01.030}}.

\bibitem[BB13]{Beer2013}
\textsc{Beer D., Burrows R.}:
\newblock {Popular Culture, Digital Archives and the New Social Life of Data}.
\newblock \emph{Theory, Culture \& Society 30}, 4 (2013), 47--71.
\newblock \href {https://doi.org/10.1177/0263276413476542} {\path{doi:10.1177/0263276413476542}}.

\bibitem[Bev13]{Bevan2013}
\textsc{Bevan A.}:
\newblock \emph{{Visualising and Quantifying the Properties of Data}}.
\newblock Cambridge University Press, 2013, p.~35–55.
\newblock \href {https://doi.org/10.1017/CBO9781139342810.005} {\path{doi:10.1017/CBO9781139342810.005}}.

\bibitem[BO61]{Beardslee1961}
\textsc{Beardslee D., O'Dowd D.}:
\newblock {The College-student Image of the Scientist}.
\newblock \emph{Science 133}, 3457 (1961), 997--1001.
\newblock \href {https://doi.org/10.1126/science.133.3457.997} {\path{doi:10.1126/science.133.3457.997}}.

\bibitem[Bod94]{Boden1994}
\textsc{Boden M.}:
\newblock Précis of the creative mind: Myths and mechanisms.
\newblock \emph{Behavioral and Brain Sciences 17}, 3 (1994), 519--531.
\newblock \href {https://doi.org/10.1017/s0140525x0003569x} {\path{doi:10.1017/s0140525x0003569x}}.

\bibitem[BS06]{Bertini2006}
\textsc{Bertini E., Santucci G.}:
\newblock {Give Chance a Chance: Modeling Density to Enhance Scatter Plot Quality Through Random Data Sampling}.
\newblock \emph{Information Visualization 5}, 2 (2006), 95--110.
\newblock \href {https://doi.org/10.1057/palgrave.ivs.9500122} {\path{doi:10.1057/palgrave.ivs.9500122}}.

\bibitem[Cle93]{cleveland1993visualizing}
\textsc{Cleveland W.~S.}:
\newblock \emph{{Visualizing Data}}.
\newblock Hobart press, 1993.

\bibitem[EB07]{Eppler2007}
\textsc{Eppler M., Burkhard R.}:
\newblock {Visual Representations in Knowledge Management: Framework and Cases}.
\newblock \emph{Journal of Knowledge Management 11}, 4 (2007), 112--122.
\newblock \href {https://doi.org/10.1108/13673270710762756} {\path{doi:10.1108/13673270710762756}}.

\bibitem[FC12]{franzblau2012graphs}
\textsc{Franzblau L.~E., Chung K.~C.}:
\newblock {Graphs, Tables, and Figures in Scientific Publications: The Good, the Bad, and How not to be the Latter}.
\newblock \emph{Journal of Hand Surgery 37}, 3 (Mar 2012), 591--596.
\newblock Epub 2012 Feb 2.
\newblock \href {https://doi.org/10.1016/j.jhsa.2011.12.041} {\path{doi:10.1016/j.jhsa.2011.12.041}}.

\bibitem[FF13]{Facione2013}
\textsc{Facione P., Facione N.}:
\newblock {Critical Thinking for Life}.
\newblock \emph{Inquiry: Critical Thinking Across the Disciplines 28}, 1 (2013), 5--25.
\newblock \href {https://doi.org/10.5840/inquiryct20132812} {\path{doi:10.5840/inquiryct20132812}}.

\bibitem[GP01]{Gershon2001}
\textsc{Gershon N., Page W.}:
\newblock {What Storytelling can do for Information Visualization}.
\newblock \emph{Communications of the ACM 44}, 8 (2001), 31--37.
\newblock \href {https://doi.org/10.1145/381641.381653} {\path{doi:10.1145/381641.381653}}.

\bibitem[HS09]{Hallstrom2009}
\textsc{Hallstrom J.~O., Soundarajan N.}:
\newblock {Reusing Patterns Through Design Refinement}.
\newblock In \emph{{Formal Foundations of Reuse and Domain Engineering}} (2009), Edwards S.~H., Kulczycki G., (Eds.), Springer Berlin Heidelberg, pp.~225--235.
\newblock \href {https://doi.org/10.1007/978-3-642-04211-9_22} {\path{doi:10.1007/978-3-642-04211-9_22}}.

\bibitem[Kei02]{Keim2002}
\textsc{Keim D.}:
\newblock {Information Visualization and Visual Data Mining}.
\newblock \emph{IEEE Transactions on Visualization and Computer Graphics 8}, 1 (2002), 1--8.
\newblock \href {https://doi.org/10.1109/2945.981847} {\path{doi:10.1109/2945.981847}}.

\bibitem[Kir19]{Kirk2019}
\textsc{Kirk A.}:
\newblock \emph{{Data Visualisation: A Handbook for Data Driven Design}}.
\newblock SAGE Publications Ltd, London, 2019.

\bibitem[MG13]{Mayorga2013}
\textsc{Mayorga A., Gleicher M.}:
\newblock {Splatterplots: Overcoming Overdraw in Scatter Plots}.
\newblock \emph{IEEE Transactions on Visualization and Computer Graphics 19}, 9 (2013), 1526--1538.
\newblock \href {https://doi.org/10.1109/tvcg.2013.65} {\path{doi:10.1109/tvcg.2013.65}}.

\bibitem[MJM{\etalchar{*}}06]{Munzner2006}
\textsc{Munzner T., Johnson C., Moorhead R., Pfister H., Rheingans P., Yoo T.}:
\newblock {NIH-NSF Visualization Research Challenges Report Summary}.
\newblock \emph{IEEE Computer Graphics and Applications 26}, 2 (2006), 20--24.
\newblock \href {https://doi.org/10.1109/mcg.2006.44} {\path{doi:10.1109/mcg.2006.44}}.

\bibitem[Moh22]{Mohanty2022}
\textsc{Mohanty S.}:
\newblock {Sound of 114 Species of Birds Till 2022}.
\newblock Kaggle, 2022.
\newblock Available at: \url{https://www.kaggle.com/datasets/soumendraprasad/sound-of-114-species-of-birds-till-2022} (Accessed: 17 May 2024).

\bibitem[Mot24]{Miebach}
\textsc{Mothes K.}:
\newblock {Nathalie Miebach Weaves Data and Anecdotes into Expansive Sculptures to Raise Awareness of the Climate Crisis}.
\newblock Colossal, 2024.
\newblock Available at: \url{https://www.thisiscolossal.com/2022/11/nathalie-miebach-data-sculptures/} (Accessed: 18 May 2024).

\bibitem[NZE08]{Nino2008}
\textsc{Ni{\~{n}}o~Zambrano R., Engelhardt Y.}:
\newblock {Diagrams for the Masses: Raising Public Awareness — from Neurath to Gapminder and Google Earth}.
\newblock In \emph{Diagrammatic Representation and Inference} (2008), Stapleton G., Howse J., Lee J., (Eds.), Springer, pp.~282--292.
\newblock \href {https://doi.org/10.1007/978-3-540-87730-1_26} {\path{doi:10.1007/978-3-540-87730-1_26}}.

\bibitem[PE01]{Paul2001}
\textsc{Paul R., Elder L.}:
\newblock \emph{{A Miniature Guide for Students on how to Study and Learn a Discipline using Critical Concepts and Tools}}.
\newblock Foundation for Critical Thinking, Dillon Beach, CA, 2001.

\bibitem[PE09]{paul2009miniature}
\textsc{Paul R., Elder L.}:
\newblock \emph{{The Miniature Guide to Critical Thinking: Concepts \& tools}}, 6th~ed.
\newblock Foundation for Critical Thinking Press, 2009.

\bibitem[Per20]{Periscopic2020}
\textsc{Periscopic}:
\newblock {Do Good with Data}, 2020.
\newblock Available at: \url{https://periscopic.com/} (Accessed: 13 April 2024).

\bibitem[RAO23]{Roberts2023}
\textsc{Roberts J.~C., Alnjar H., Owen Aron;~Ritsos P.~D.}:
\newblock {A method for Critical and Creative Visualisation Design-Thinking}.
\newblock In \emph{IEEE VIS 2023 Posters: Visualization and Visual Analytics} (2023).
\newblock URL: \url{https://virtual.ieeevis.org/year/2023/poster_v-vis-posters-1051.html}.

\bibitem[RHR16]{RobertsHeadleandRitsos16}
\textsc{Roberts J.~C., Headleand C.~J., Ritsos P.~D.}:
\newblock {Sketching Designs Using the Five Design-Sheet Methodology}.
\newblock \emph{IEEE Transactions on Visualization and Computer Graphics 22}, 1 (2016), 419--428.
\newblock \href {https://doi.org/10.1109/TVCG.2015.2467271} {\path{doi:10.1109/TVCG.2015.2467271}}.

\bibitem[SH10]{Segel2010}
\textsc{Segel E., Heer J.}:
\newblock {Narrative visualization: Telling Stories with Data}.
\newblock \emph{IEEE Transactions on Visualization and Computer Graphics 16}, 6 (2010), 1139--1148.
\newblock \href {https://doi.org/10.1109/tvcg.2010.179} {\path{doi:10.1109/tvcg.2010.179}}.

\bibitem[Shand]{Shaw}
\textsc{Shaw W.}:
\newblock {Synaesthetic Sanctuary}.
\newblock William Shaw, n.d.
\newblock Available at: \url{https://www.williamshawdesign.com/synaesthetic-sanctuary} (Accessed: 18 May 2024).

\bibitem[Shn96]{Shneiderman1996}
\textsc{Shneiderman B.}:
\newblock {The Eyes Have it: A Task by Data Type Taxonomy for Information Visualizations}.
\newblock In \emph{Proceedings 1996 IEEE Symposium on Visual Languages} (1996).
\newblock \href {https://doi.org/10.1109/vl.1996.545307} {\path{doi:10.1109/vl.1996.545307}}.

\bibitem[SMT{\etalchar{*}}10]{Schwamborn2010}
\textsc{Schwamborn A., Mayer R.~E., Thillmann H., Leopold C., Leutner D.}:
\newblock {Drawing as a Generative Activity and Drawing as a Prognostic Activity}.
\newblock \emph{Journal of Educational Psychology 102}, 4 (2010), 872--879.
\newblock \href {https://doi.org/10.1037/a0019640} {\path{doi:10.1037/a0019640}}.

\bibitem[SSB{\etalchar{*}}15]{Shao2015}
\textsc{Shao L., Schleicher T., Behrisch M., Schreck T., Sipiran I., Keim D.~A.}:
\newblock {Guiding the Exploration of Scatter Plot Data Using Motif-based Interest Measures}.
\newblock In \emph{{Big Data Visual Analytics (BDVA)}} (2015).
\newblock \href {https://doi.org/10.1109/bdva.2015.7314294} {\path{doi:10.1109/bdva.2015.7314294}}.

\bibitem[TS07]{Tu2007}
\textsc{Tu Y., Shen H.-W.}:
\newblock {Visualizing Changes of Hierarchical Data using Treemaps}.
\newblock \emph{IEEE Transactions on Visualization and Computer Graphics 13}, 6 (2007), 1286--1293.
\newblock \href {https://doi.org/10.1109/tvcg.2007.70529} {\path{doi:10.1109/tvcg.2007.70529}}.

\bibitem[WHN{\etalchar{*}}16]{WangEtal2016}
\textsc{Wang W.~B., Huang M.~L., Nguyen Q.~V., Huang W., Zhang K., Huang T.-H.}:
\newblock {Enabling Decision Trend Analysis with Interactive Scatter Plot Matrices Visualization}.
\newblock \emph{Journal of Visual Languages \& Computing 33} (2016), 13--23.
\newblock SI:IVIMLA.
\newblock \href {https://doi.org/10.1016/j.jvlc.2015.11.002} {\path{doi:10.1016/j.jvlc.2015.11.002}}.

\end{thebibliography}





\end{document}